\documentclass[final]{IEEEtran}
\usepackage{blindtext, graphicx}
\usepackage{mymacros}

\begin{document}
%

\title{Coverage and Spectral Efficiency of Indoor mmWave Networks with Ceiling-Mounted Access Points}




%
\author{
\IEEEauthorblockN{Fadhil Firyaguna, Jacek Kibi\l{}da, Carlo Galiotto, Nicola Marchetti}

\IEEEauthorblockA{CONNECT Centre, Trinity College Dublin, Ireland \\
\{firyaguf, kibildj, galiotc, nicola.marchetti\}@tcd.ie}
}


\pagestyle{empty}

\maketitle
\thispagestyle{empty}

\begin{abstract}
    Provisioning of high throughput millimetre-wave signal to indoor areas that potentially serve a large number of users, such as transportation hubs or convention centres, will require dedicated indoor millimetre-wave access point deployments.
    In this article, we study dense deployments of millimetre-wave access points mounted on the ceiling, and illuminating selected spots on the ground with the use of fixed directional antennas. In this setup, the main factor limiting signal propagation are blockages by human bodies.
    We evaluate our system under a number of scenarios that take into account beamwidth of the main-lobe, access point density, and positioning of the mobile device with respect to the user's body.
    We find that both coverage and area spectral efficiency curves exhibit non-trivial behaviour which can be classified into four regions related to the selection of access point density, beamwidth, and height values.
    Furthermore, we observe a trade-off in beamwidth design, as the optimal beamwidth maximizes either coverage or area spectral efficiency, but not both.
    Finally, when we consider different body shadowing scenarios, our network design optimizes coverage or area spectral efficiency performance towards either devices held in hand or worn directly against the body, as each of the scenarios requires mutually exclusive settings of access point density and beamwidth.
\end{abstract}


%


\section{Introduction}

According to the taxonomy provided by ETSI \cite{ETSI_2015}, \ac{mmWave} spectrum spans frequencies from \unit[50]{GHz} to \unit[300]{GHz}. Systems that could provide reliable communications over these frequencies attract great attention as the said frequencies offer much wider bandwidths at shorter wavelengths, in comparison to micro-wave frequencies. While wider bandwidth may be directly translated to increased link throughput, the shorter wavelength may allow networks to take greater advantage of techniques that increase power concentration at the receiver and spatial separation between the transmitters, resulting in capacity gains. Coarse estimates provided by ETSI show that, even with a single-antenna, a \unit[500]{MHz} 16-QAM \ac{mmWave} link may achieve over \unit[1]{Gbps} of throughput. This signifies that, if \ac{mmWave} systems are shown to be technically and commercially feasible, they could be used to address the capacity objectives of 5G.

Yet, cellular systems that utilize \ac{mmWave} frequencies will likely be providing coverage that is confined to streets and, more generally, outdoor areas only, as \ac{mmWave} signals do not propagate well through physical objects \cite{PiKhan_2011}. This creates a situation in which an independent tier of \ac{mmWave} \acp{AP} would be required to ensure even basic coverage over indoor areas that serve potentially large number of end users, such as concert halls, transportation hubs, or convention centres. 

State-of-the-art literature on \ac{mmWave} communications has shown that \ac{mmWave} deployments can be a source of high bit-rate signal for indoor users (see, for example, \cite{dong2012link,venugopal2016device}) but, as we will discuss in the following section, it has not provided much in the way of network-level design and radio access infrastructure deployment. In this paper we close this gap by studying the performance effects of deployment densification of ceiling-mounted \ac{mmWave} access points with highly directional antennas over a confined area.

In our scenario \ac{mmWave} access points are mounted on the ceiling or walls to form a grid-like pattern and set to illuminate selected spots on the ground. In this case, and given the significantly shorter distances between the \acp{AP} and \acp{UE} than in an outdoor scenario, the main factor limiting signal propagation are blockages by human bodies, which have been shown to introduce as much as \unit[40]{dB} of attenuation to the propagating \ac{mmWave} signal (see, for example, \cite{collonge2004influence,rajagopal2012channel}). Moreover, the potential lack of fixed physical obstructions such as inner-walls may result in interference between adjacent \acp{AP}, despite the usage of directional transmissions. Effectively, deploying such networks requires understanding of the relationship between basic design parameters such as \ac{AP} density, main lobe width, or transmit power and the propagation features of \ac{mmWave} signals.

What we find is that both the coverage and \ac{ASE} curves display non-trivial behaviour which can be classified into four regions related to the selection of \ac{AP} density, beamwidth and height values. Furthermore, we find that there is a trade-off in beamwidth design, as the optimal beamwidth maximizes either coverage or \ac{ASE}, but not both. This trade-off gets more complicated when we consider an indoor \ac{mmWave} scenario where human body introduces significant attenuation to the propagating signal which cannot be fully compensated for with handovers (as we show in the analysis of the cell association policy). To better understand this we compare the coverage and \ac{ASE} for two scenarios of human body shadowing: a \ac{UE} operated in front of the user (\yale{UE in hand}) and a \ac{UE} located in the pocket or carried as a wearable (\yale{UE in pocket}). 
In the former scenario, the peak coverage requires that we use denser deployment and smaller beamwidths, which is shown to be beneficial also to the achieved \ac{ASE}.\footnote{The corresponding \ac{ASE} achieved with the optimal beamwidth for coverage.}
In the latter scenario, the peak coverage requires that we use lower deployment densities and larger beamwidths, although this  configuration is not optimal for the achieved \ac{ASE}.

In what follows we provide an overview of the related literature, a description of our system model, and an in-depth analysis of the numerical results obtained, with lessons learnt on the design of dense indoor \ac{mmWave} networks.

\section{Related Work}

Our goal is to study the performance effects of deployment densification of ceiling-mounted \ac{mmWave} access points with highly directional antennas over a confined area. While state-of-the-art literature has not addressed this topic directly, there are various other well-studied subjects, such as network densification, which provide us with relevant conclusions.  

Network densification is key to increasing the capacity of conventional mobile networks, as spectrum designated for cellular communications in microwave frequencies is relatively scarce. In the \ac{mmWave} frequencies, where spectrum is in abundance but adverse propagation conditions limit the signal penetration, network densification may be used to shorten the physical distance between the transmitters and receivers, ramping up the signal level at the receivers' input. Indeed, dense \ac{mmWave} networks have been shown to be an attractive deployment option for outdoor urban areas \cite{bai2013coverage,bai2013coverageindense,bai2014analysis,bai2015coverage}. In \cite{bai2013coverageindense} it has been shown that optimal operation of a wide-area \ac{mmWave} system requires a deployment that is dense enough to ensure line-of-sight conditions from at least a few transmitters. Lower density deployments result in significantly lower performance due to non-line-of-sight operation, while higher density deployments lead to an increase in interference which deteriorates the system's performance. Wide-area cellular systems based on \ac{mmWave} frequency bands also require extensive indoor deployments as \ac{mmWave} signals do not penetrate well majority of materials \cite{PiKhan_2011}.

In fact, as early as 2011, WiGig in cooperation with IEEE 802.11, proposed a PHY/MAC layer that was dedicated towards wireless local area operation in \ac{mmWave} frequency bands (see \cite{Hansen_2011}). The proposed technology was integrated with WiFi standards operating in microwave frequency bands allowing for a graceful fallback to microwave spectrum operation when needed (see \cite{Hansen_2011}). Number of research studies have confirmed the technology to be capable of delivering \unit[]{Gbit/s} link throughputs over a range of up to 10 metres in line-of-sight conditions (see, for example, \cite{ZhuDoufexiKocak_2011}). However, the network-level performance of \ac{mmWave} indoor deployments, such as WiGig (or 802.11ad as it is currently known) remains largely unknown.

In a mmWave indoor scenario, characterized by much smaller distances between access points and users, the main factor limiting deployment options are blockages by physical objects such as human bodies. Human body blockage was shown to cause severe signal blockages (as high as \unit[40]{dB}) that reduce the spectral efficiency gains obtained from operation over larger bandwidths available in \ac{mmWave} frequencies (see \cite{lu2012modeling,rappaport2013millimeter}). In small enclosed areas this detrimental effect of body-related shadowing can be at least partially mitigated by application of reflective materials to vertical surfaces and usage of signal relays (see \cite{leong2004robust,dong2012link}). However, in large open indoor areas these may not necessarily be available and, moreover, lack of fixed physical obstructions such as walls may actually lead to significant interference between adjacent access points requiring that the \ac{mmWave} link performance is considered from network-level perspective. In \cite{wells2009faster} the trade-off between the received signal strength and the probability of blockage when deciding on the transmit antenna height is reported. Simply shortening the distance to the receiver by lowering antenna heights (thereby reducing distance-dependent pathloss) yields a greater chance of the signal being blocked by the human body, especially in crowded areas~\cite{dong2012link,venugopal2015analysis}. This trade-off can be exploited to study optimal altitude for signal-providing low altitude platforms~\cite{al2014optimal}, such as quadcopters, or balloons, or urban outdoor cellular deployments with blockages from human bodies~\cite{gapeyenko2016analysis}. Furthermore, as it was shown in \cite{bai2014analysis}, increase in human body blockage loss increases \yale{coverage inequality} in the system, as receivers with poor coverage observe a further reduction to coverage, while good coverage receivers see their coverage being improved. In this scenario, whether you observe a drop or increase in coverage depends on whether the human body is shadowing more the serving transmitter (poor coverage users) or the interferers (good coverage users). In \cite{venugopal2016device}, which studies device-to-device indoor \ac{mmWave} communications scenario, it is shown that, under the assumption of a random direction of the interferer's main-lobe, highly directional beams will be required to maintain \unit[]{Gbit/s} links in crowded indoor areas.  

Despite these detailed insights on the impact of heights, fixed beamwidths and body blockage on the performance of both conventional and \ac{mmWave} links, inspected both in isolation and in the network context, still little is known about the coverage and \ac{ASE} trade-offs in densification of ceiling-mounted \ac{mmWave} access points. In the following, building on the state-of-the-art literature for \ac{mmWave} network modelling \cite{andrews2017modeling}, we setup a system model that allows us to inspect the trade-offs between peak coverage and \ac{ASE} given variety of blockage scenarios and cell association strategies.

\section{System Model}
\label{sec:systemmodel}
The considered environment is an indoor confined area where the main obstacles for the mmWave signal propagation are human bodies, i.e., we assume a scenario with no corridors or walls such as theatres and convention centre halls. The \acp{AP} are deployed on a hexagonal grid throughout the indoor venue, and they are installed on the ceiling at a height $h_\mathrm{AP}$ above the \ac{UE} level, with fixed directional antennas illuminating the floor below. 
We consider a \ac{UE} randomly located in the cell at the centre of the venue. The \ac{UE} is associated with the serving \ac{AP} for the downlink transmission by a given cell association strategy defined in Section~\ref{sec:simulationsetup}.

\subsection{Directivity Gain}
\begin{figure}
    
    \begin{subfigure}{.5\linewidth}
        \centering
        \includegraphics[scale=.225]{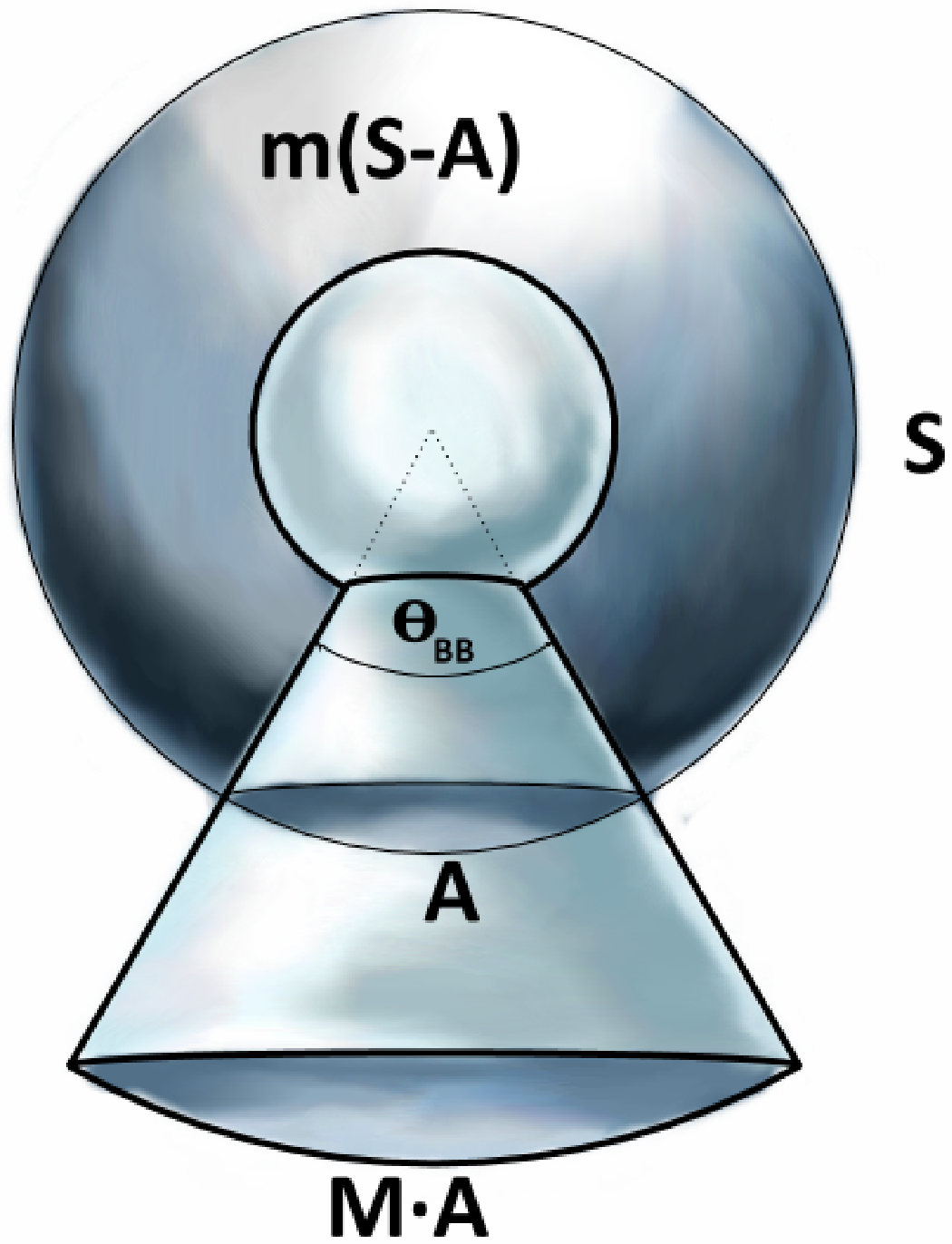}
        \caption{}
        \label{fig:conebulbmodel}
    \end{subfigure}
    \begin{subfigure}{.5\linewidth}
        \centering
        \includegraphics[clip,scale=.45]{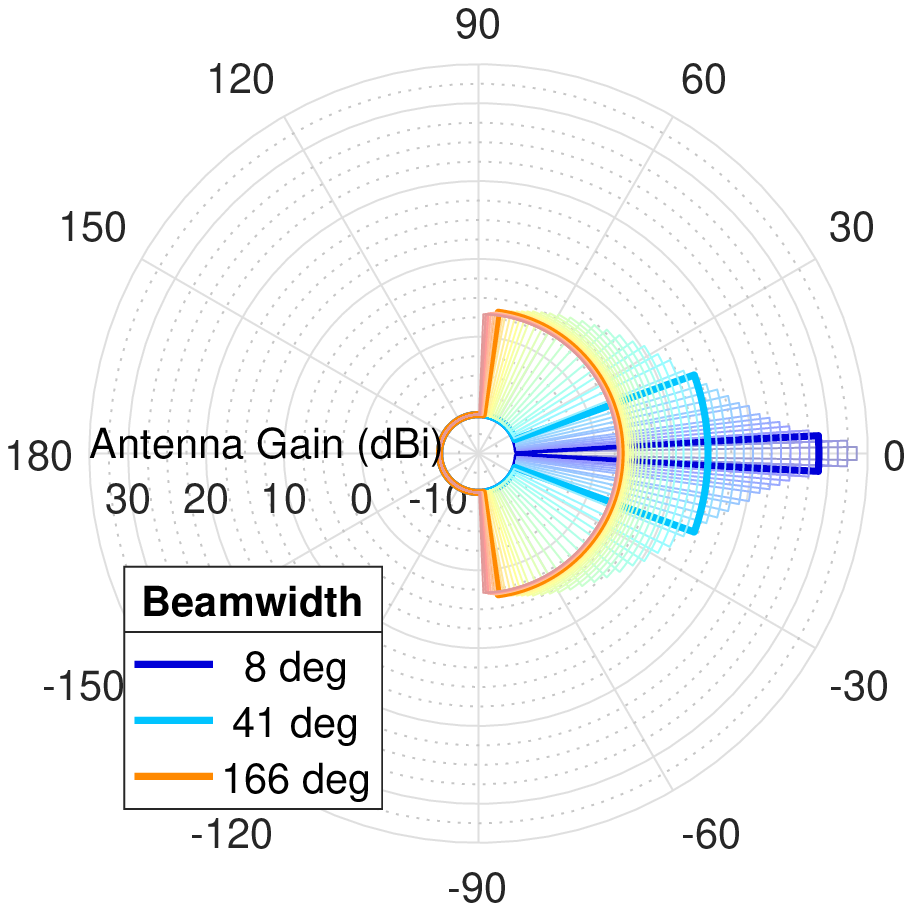}
        \caption{}
        \label{fig:antennapattern}
    \end{subfigure}
    \caption{(a) Cone-bulb model approximation of antenna directivity gain patterns. The cone represents the main-lobe and the bulb (inner sphere) represents the side-lobe. The cap area $A$ is ``amplified'' by a factor of $M$, while the resulting bulb area $S-A$ is ``shrunk'' by a factor of $m$. (b) Antenna directivity gain pattern for different beamwidths. }
\end{figure}

We assume \acp{AP} utilize directional transmission, while the \acp{UE} utilize omnidirectional reception. As in \cite{ramanathan2001performance}, the antenna pattern is approximated by a cone of uniform gain representing the main-lobe attached to a single ``bulb" representing the side-lobe, as illustrated in Fig.~\ref{fig:conebulbmodel}, where $M$ is the main-lobe directivity gain, $m$ is the side-lobe gain, $\theta_\mathrm{BW}$ is the beamwidth of the main-lobe, $A$ is the area of the spherical cap, and $S$ is the surface area of the sphere.
The directivity gains are then a function of the beamwidth, normalized over a given spherical surface as in:
\begin{equation}
    M \cdot \frac{A}{S} + m \cdot \frac{S-A}{S} = 1,
    \label{eq:beamformgain}
\end{equation}
where the area of the cap is given by $A = 2 \pi r^2 (1-\cos \frac{\theta_\mathrm{BW}}{2})$, and the sphere surface area is $S = 4 \pi r^2$. Thus, fixing the side-lobe gain $m$, we can calculate the main-lobe gain as a function of the beamwidth:
\begin{equation}
    M(\theta_\mathrm{BW},m) = \frac{2 - m ( 1 + \cos \frac{\theta_\mathrm{BW}}{2} )}{1 - \cos \frac{\theta_\mathrm{BW}}{2}}.
    \label{eq:mainlobegain}
\end{equation}


The \ac{UE} receives maximum directivity gain $M$ of an \ac{AP} when the \ac{UE} is positioned in the illumination area of the main-lobe of that \ac{AP}, i.e., the \ac{UE} is inside the projected circle of the main-lobe of radius:
\begin{equation}
    r_M = h_\mathrm{AP} \cdot \tan\frac{\theta_\mathrm{BW}}{2},
    \label{eq:mainloberadius}
\end{equation}
as illustrated in Fig.~\ref{fig:topheadblock}.
Otherwise, the directivity gain is the \ac{AP} side-lobe gain $m$ (as shown in Fig.~\ref{fig:antennapattern}).

\subsection{Self-Body Blockage}
\begin{figure}
   \centering 
    \begin{subfigure}{.6\linewidth}
        \centering
        \includegraphics[scale=.4]{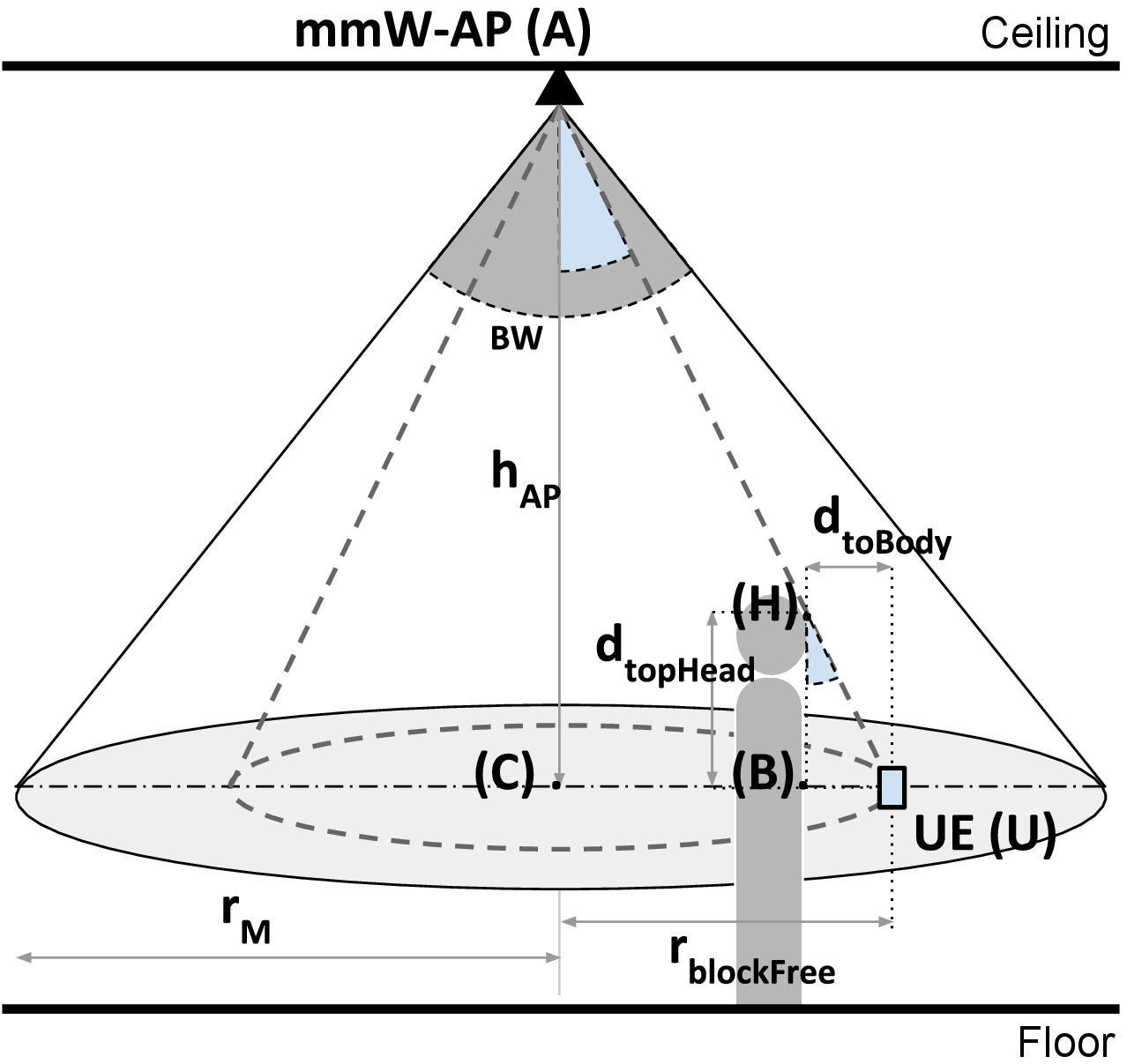}
        \caption{Side view.}
        \label{fig:topheadblock}
    \end{subfigure}%
    \begin{subfigure}{.4\linewidth}
        \centering
        \includegraphics[scale=.4]{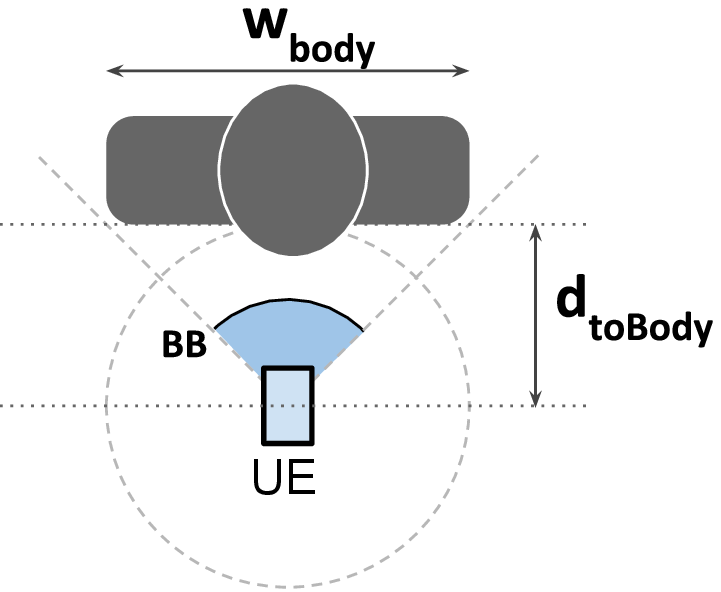}
        \caption{Top view.}
        \label{fig:bodyblock}
    \end{subfigure}
    \caption{Self-body blockage model. Side view (a): vertical obstruction by the user body may happen when the \ac{UE} is beyond the radius $r_\mathrm{blockFree}$. User is holding the \ac{UE} at a distance of $d_\mathrm{toBody}$ in front of the body, and at a distance of $d_\mathrm{topHead}$ to the top of the head level.  Top view (b): the user body horizontally blocks a region around the \ac{UE} defined by the angle $\theta_\mathrm{BB}$.}
    \label{fig:selfbodyblock}
\end{figure}

In our scenario, the only source of blockage is a human body. Body blockage can cause up to \unit[40]{dB} of attenuation to the penetrating signal \cite{lu2012modeling,bai2014analysis,rajagopal2012channel,collonge2004influence}. 
The main factor that describes the extent to which human body shadows signals to/from the \ac{UE} is the \ac{UE}'s position with respect to the body. This position is determined by two parameters: $d_\mathrm{toBody}$ and $d_\mathrm{topHead}$ (as shown in Fig.~\ref{fig:topheadblock}). 
The first one determines the distance to the body and how wide the signal obstruction is, e.g., zero distance could represent a scenario where the device is held in a pocket and the body obstructs half of the field of view, while a distance of \unit[30]{cm} could represent a scenario where a user is browsing the Internet and the body obstructs a narrower area. The second parameter determines the amount of body obstruction in the vertical dimension.
Given the body blockage and our ceiling-mounted deployments, we can construct a model of user device shadowing as depicted in Fig.~\ref{fig:topheadblock}. From the geometry of the model, we define $r_\mathrm{blockFree}$ as the radius of the \textit{self-block free zone}:
\begin{equation}
    r_\mathrm{blockFree} = h_\mathrm{AP} \cdot \frac{d_\mathrm{toBody}}{d_\mathrm{topHead}},
    \label{eq:blockfreeradius}
\end{equation}
where the \acp{UE} inside this zone are never obstructed by the user body, while the \acp{UE} outside of it are obstructed whenever the user body is between the \ac{UE} and the \ac{AP}. 
Now, assuming uniform body orientation, the probability of a user body obstructing the \ac{AP}'s signal (self-block probability) is:
\begin{equation}
    P_\mathrm{BB} = \arctan\left(\frac{w_\mathrm{body}}{2 \cdot d_\mathrm{toBody}}\right) / \pi.
    \label{eq:prob.bodyblock}
\end{equation}

\subsection{Signal-to-Interference-Noise Ratio}

In this work, we consider the following path loss model:

\begin{equation}
    L(d,h_\mathrm{AP}) = L_0 \cdot R(d,h_\mathrm{AP})^{-\alpha},
    \label{eq:los.pathloss}
\end{equation}
where $L_0$ is the path loss at 1 metre distance under free space propagation, $\alpha$ is the attenuation exponent, $d$ is the projection of the distance on the horizontal plane (2D-distance) from the cell centre to the \ac{UE}, and $R(d,h_\mathrm{AP})$ is the Euclidean distance from the \ac{AP} to the \ac{UE}. 

Based on the assumptions made above, we can express the SINR at a \ac{UE} as:
\begin{equation}
    \mathrm{SINR} = \frac{G_i \cdot L( d_i, h_\mathrm{AP} ) \cdot B_i}{ N_0 / P_\mathrm{TX} + \sum_{j \in \Dc \setminus \{i\}} G_j \cdot L( d_j, h_\mathrm{AP} ) \cdot B_j},
\end{equation}
where $\Dc$ represents the set of all \acp{AP} in the system, $d_i$ is the distance to the serving \ac{AP} $i\in\Dc$, $G_i \in \{m,M\}$ is the directivity gain of \ac{AP} $i$, $B_i \in \{L_\mathrm{body},1\}$ is the body attenuation for the link between the reference user and \ac{AP} $i$ ($L_\mathrm{body}$ is the attenuation loss produced by the body), $N_0$ is the thermal noise power, and $P_\mathrm{TX}$ is the transmit power. Note that $G_i$ and $B_i$ are random variables whose distributions are functions of the system parameters ($\theta_\mathrm{BW}$, $h_\mathrm{AP}$, $d_\mathrm{toBody}$, $d_\mathrm{topHead}$, $w_\mathrm{body}$), and distance $d_i$.

In our scenario, we assume there are no physical obstructions to the propagating signal other than the user's body; in addition, we consider the reflections from ceiling and ground to be negligible,
which may be considered a reasonable modelling assumption since,
as reported in \cite{yiu2009empirical,genc2010robust},
several materials used for ceiling and flooring surfaces produce a significant attenuation in the reflected signal.

\section{Numerical Results}

\subsection{Simulation Setup}
\label{sec:simulationsetup}
\begin{figure}
    \centering
    \includegraphics[scale=.52]{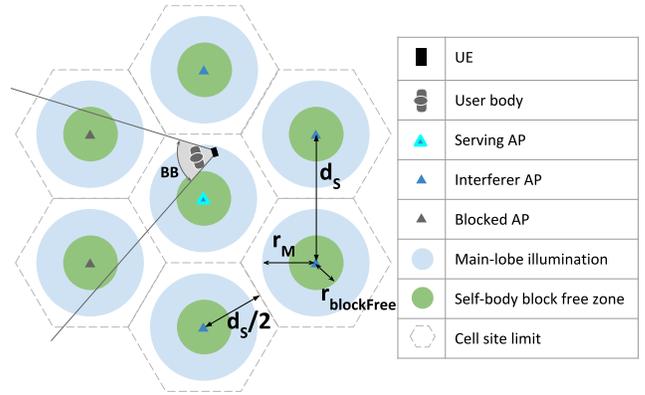}
    \caption{Snapshot from simulations illustrating the system model. The \acp{AP} are distributed according to a hexagonal cell pattern with an inter-site distance $d_s$. User body is blocking the signal from the gray-colored \acp{AP}. The \ac{UE} is illuminated (light-blue area) by the serving \ac{AP} and is not within the self-block free zone (green area). Note that in a very dense topology, where $d_s$ could be as small as $r_\mathrm{blockFree}$, the site area could correspond to self-block free zone.}
    \label{fig:topology}
\end{figure}

We model our scenario by placing \acp{AP} in the centres of a hexagonal cell pattern laid over a \unit[400 $\times$ 400]{m$^2$} area, as exemplified in Fig.~\ref{fig:topology}. This specific choice of the area size allows us to mitigate the edge effect, and to explore the system behaviour for longer inter-site distances. The side-lobe gain is fixed at \unit[$-10$]{dB}, and the main-lobe gain varies with the beamwidth according to Eq.~(\ref{eq:mainlobegain}). We evaluate the system for a fixed \ac{AP} height $h_\mathrm{AP}$~=~\unit[10]{m}. Note that, changing $h_\mathrm{AP}$ has essentially the same impact on the performance as changing the beamwidth, since both $h_\mathrm{AP}$ and $\theta_{BW}$ determine the main-lobe illumination area; as a matter of fact, when testing our system for other height values of interest, we observed no significant deviations from our conclusions. We set the attenuation exponent as $\alpha=2$, transmit power as \unit[20]{dBm}, bandwidth as \unit[100]{MHz}, carrier frequency as \unit[60]{GHz}, noise figure as \unit[9]{dB}, we consider no small-scale fading, and we set the body parameters $w_\mathrm{body}$ as \unit[40]{cm} and $d_\mathrm{topHead}$ as \unit[40]{cm}. We set the parameter $d_\mathrm{toBody}$ to define different blockage scenarios: $d_\mathrm{toBody}$~=~\unit[30]{cm} represents a scenario of a user operating the \ac{UE} with the \textit{hand} (\ac{UE} in hand), and $d_\mathrm{toBody}$~=~\unit[0]{cm},\footnote{In this scenario, $r_{blockFree}$ equals 0, according to Eq.~\ref{eq:blockfreeradius}, and there is no self-block free zone.} represents a scenario of the \ac{UE} held in \textit{pocket} (\ac{UE} in pocket). The \ac{UE} is associated with an \ac{AP} that provides it with a downlink signal according to either the shortest 3D Euclidean distance (\textit{minimum distance association}) or the strongest received signal power (\textit{maximum received power association}). The simulation  source code is available on-line (see Appendix~\ref{ap:simulationcodes}).

\subsection{Coverage and Area Spectral Efficiency Profile}
\label{sec:profile}

\begin{figure}[ht!]
    \centering
    \begin{subfigure}{\linewidth}
    \includegraphics[width=\linewidth]{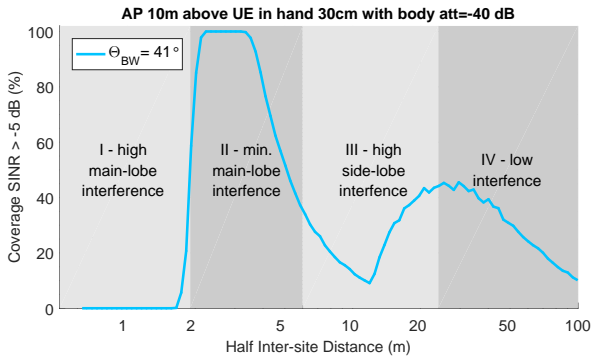}
    \caption{Regions delimited by gray rectangles represents the non-trivial behaviour of coverage in the selected \ac{AP} density.}
    \label{fig:cov-5_ap10_hand_mind_profile}
    \end{subfigure}
    \hfill
    
    \begin{subfigure}{\linewidth}
    \includegraphics[width=\linewidth]{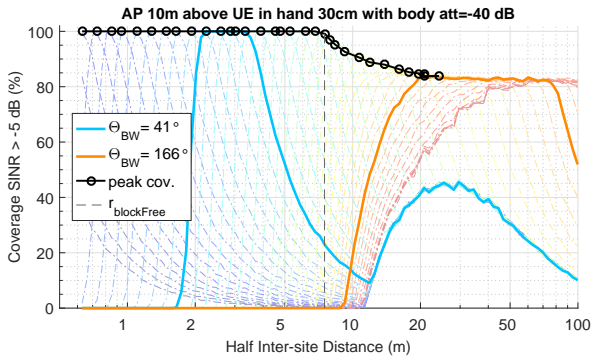}
    \caption{Coverage for SINR threshold above \unit[-5]{dB}.}
    \label{fig:cov-5_ap10_hand_mind}
    \end{subfigure}
    \hfill
    
    \begin{subfigure}{\linewidth}
    \includegraphics[width=\linewidth]{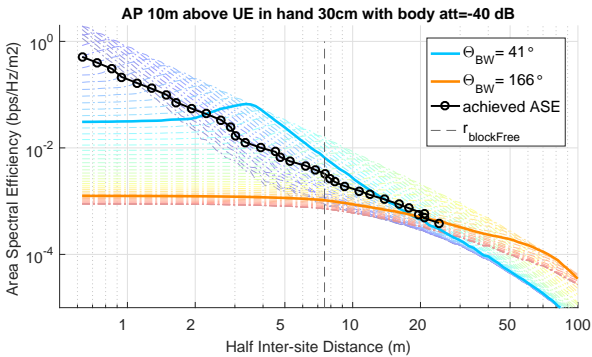}
    \caption{Area spectral efficiency.}
    \label{fig:ase_ap10_hand_mind}
    \end{subfigure}
    
    \caption{Coverage and ASE for different beamwidths (light and dotted lines) when the user holds the \ac{UE} in hand and the \acp{AP} are mounted \unit[10]{m} above the \ac{UE}. The results for beamwidths of \unit[41]{$^\circ$} and \unit[166]{$^\circ$} are pictured in solid lines. The black line in (a) is the interpolation of the maximum coverage achieved for a given beamwidth and inter-site distance. The black line in (b) is the ASE achieved from choosing a given pair beamwidth/inter-site distance. The gray vertical dashed line is the radius of the self-block free zone.}
    \label{fig:cov-5_ase_ap10_hand_mind}
\end{figure}

In this subsection, we evaluate the effect of the inter-site distance (network density) and beamwidth on coverage and area spectral efficiency (\ac{ASE}) of a mmWave indoor network with ceiling-mounted \acp{AP}. The SINR coverage is defined as the probability that the SINR at the receiver is larger than some threshold $T$, i.e., $P[\mathrm{SINR}>T]$, while the \ac{ASE} is the spectral efficiency $\log(1+\mathrm{SINR})$ averaged over all realizations, and divided by the cell area. The results for coverage and ASE are shown in Fig.~\ref{fig:cov-5_ap10_hand_mind} and \ref{fig:ase_ap10_hand_mind}, respectively; for now, we focus on the minimum distance cell association case. 

Our investigation reveals that, in the ceiling-mounted \ac{AP} setup, the SINR coverage presents a non-trivial behaviour which can be classified into four regions, as illustrated in Fig.~\ref{fig:cov-5_ap10_hand_mind_profile}. These appear as we change the inter-site distance while keeping the beamwidth fixed: 
\begin{inparaenum}[(i)] 
\item \textbf{high main-lobe interference}: at high \ac{AP} density (short $d_S$), the beam is too large and causes substantial overlaps among adjacent cells, which results in high interference and, thus, low coverage; 
\item \textbf{minimum main-lobe interference}: the main-lobe illuminates the entire cell with minimum interference to neighbouring cells, yielding high coverage; from this point on, as we move towards a sparser deployment, the cell size becomes larger than the illuminated area and the coverage is inevitably reduced; 
\item \textbf{high side-lobe interference}: at intermediate \ac{AP} densities, the coverage is very low due to the lack of main-lobe illumination by the serving \ac{AP} and due to high neighbour side-lobe interference; however, this interference decreases as the deployment gets sparser, leading to increased coverage; 
\item \textbf{low interference}: in low \ac{AP} density (large $d_S$), the beam is so small that it becomes negligible; therefore, the only signal that can be picked-up by the majority of users comes from the side-lobe and is thus weak enough for the noise to dominate the SINR term.
\end{inparaenum}

Based on these results, it is clear that for each cell size (or each \ac{AP} density) there is an optimal design of beamwidth that leads to a peak coverage, which is depicted by the black line in Fig.~\ref{fig:cov-5_ap10_hand_mind}. For example, the half inter-site distance $d_S / 2$~\unit[$\approx $~3.4]{m} corresponds to peak coverage when $\theta_\mathrm{BW}$~=~\unit[41]{$^\circ$} (which is equivalent to the main-lobe radius $r_M$~=~\unit[3.7]{m}). It should be noted that the optimal beamwidth for \unit[$-5$]{dB} coverage does not optimize the \ac{ASE}. As we see from the black line in Fig.~\ref{fig:ase_ap10_hand_mind}, the achieved \ac{ASE} is lower than the maximum achievable for a given inter-site distance.  
A more detailed discussion of this trade-off between coverage and \ac{ASE} is presented in Section~\ref{sec:tradeoff}.

With reference to  Fig.~\ref{fig:cov-5_ap10_hand_mind}, the fact that we observe high peak coverage at high AP densities and relatively lower coverage at lower  AP densities depends on whether the cell size is smaller than the self-block free zone; one should note that when the cell size is smaller than the self-block free zone (on the left of the dashed line in Fig.~\ref{fig:cov-5_ap10_hand_mind}), all \acp{UE} are free from self-blockage, leading to high peak coverage. On the other hand, when the cell is bigger than the self-block free zone (on the right of the dashed line), there are some \acp{UE} outside the self-block free zone that will be blocked by the body with probability $P_\mathrm{BB}$, according to Eq.~\ref{eq:prob.bodyblock}. These \acp{UE} will have their SINR degraded by the body attenuation, increasing the number of \acp{UE} whose SINR is below the threshold. Hence, the coverage will decrease proportionally to the number of blocked \acp{UE}.


\subsection{Cell Association and Body Blockage}
\label{sec:association}

In this subsection, we investigate the impact of \ac{UE}-to-\ac{AP} association on peak coverage. To that end, we compare two different association strategies, namely \textit{minimum distance} and \textit{maximum received power} (as defined in Section~\ref{sec:simulationsetup}) and we consider two different scenarios of interest, i.e., \ac{UE} in hand --- which represents a typical device usage --- and \ac{UE} in pocket  --- which represents a severe blockage scenario. 
\begin{figure}
    \centering
    \begin{subfigure}{\linewidth}
    \includegraphics[width=\linewidth]{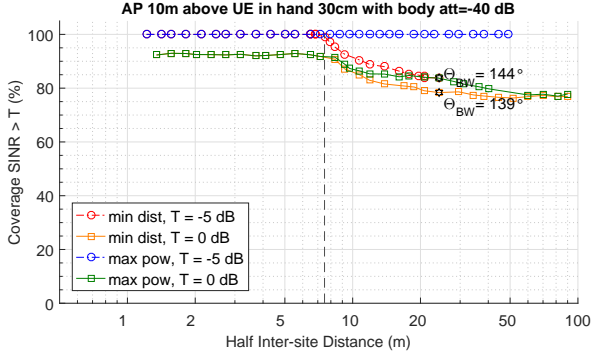}
    \caption{UE in hand.}
    \label{fig:cov_threshold_ap10_hand}
    \end{subfigure}
    \hfill
    
    \begin{subfigure}{\linewidth}
    \includegraphics[width=\linewidth]{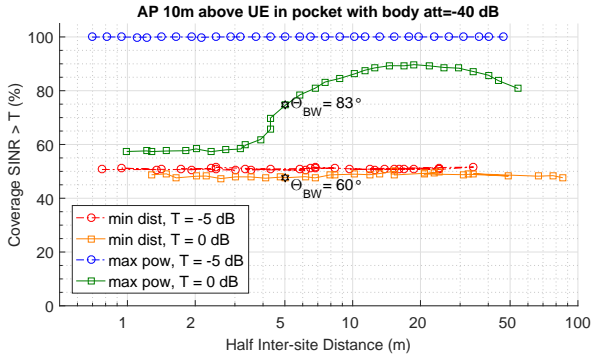}
    \caption{UE in pocket.}
    \label{fig:cov_threshold_ap10_pocket}
    \end{subfigure}
    \caption{Comparison of peak coverage for different SINR thresholds for two cell association strategies: minimum distance (red/orange curves) and maximum received power (blue/green curves). Optimal $\theta_\mathrm{BW}$, for a given $d_S$ (black marks), is generally larger when using maximum received power association.}
    \label{fig:cov_threshold_association}
\end{figure}
The corresponding results are shown in Fig.~\ref{fig:cov_threshold_ap10_hand} and \ref{fig:cov_threshold_ap10_pocket}, respectively. First, it is important to remark that the maximum received power association strategy leads to larger optimal beamwidths compared to the minimum distance strategy, meaning that it yields a cellular deployment with larger overlaps between adjacent \ac{AP} main-lobes. For example, in Fig.~\ref{fig:cov_threshold_ap10_pocket}, for $d_S/2$ = \unit[5.02]{m} and $T$ = \unit[0]{dB}, the maximum received power association (green line) leads to the optimal beamwidth of \unit[83]{$^\circ$} (UE in pocket), while the minimum distance association (orange line) leads to the optimal beamwidth of \unit[60]{$^\circ$}. Second, our results show that, as expected, maximum power association strategy generally improves the coverage of the network. 

In addition, there are a few minor observations that can be made. 
First, in the \ac{UE} in pocket scenario, with the minimum distance strategy, the coverage achieves approximately 50\%. 
This is because in this scenario, the probability of blockage $P_\mathrm{BB}$ is 50\%, which means that half of the users will block the signal to their \ac{UE}, attenuating the signal by \unit[40]{dB}. Since, in the minimum distance strategy, the blocked \acp{UE} will not associate with another \ac{AP}, those users will have a poor SINR, leading to approximately 50\% of the users not being covered. 
Second, the coverage can be improved with the maximum received power strategy because those 50\% of users will associate with another \ac{AP} and will have a better SINR. Nonetheless, this improvement depends on the SINR threshold. For example, with $T$~=~\unit[$-5$]{dB}, we have 100\% coverage at any \ac{AP} density (see blue curve in  Fig.~\ref{fig:cov_threshold_ap10_pocket}), whereas with $T$~=~\unit[0]{dB}, the coverage is lower than in the former case (up to 90\%)\footnote{Even with the maximum received power association, the \unit[0]{dB} threshold coverage does not reach 100\% in any of the body shadowing scenarios we have taken into account (an observation which coincides with the one made in \cite{bai2014analysis}).} and presents a drop for high \ac{AP} densities (as we can see from the green curve in Fig.~\ref{fig:cov_threshold_ap10_pocket} for half inter-site distances below \unit[5]{m}).
The reason for this particular behaviour of the green curve in Fig.~\ref{fig:cov_threshold_ap10_pocket} is the following. 
At high \ac{AP} densities, the SINR of the majority of the \acp{UE} is low (in particular, lower than the threshold) because of the strong interference from neighbouring \acp{AP}, which is caused by the short distances between these \acp{AP} and the \ac{UE}, and by high directivity gains (i.e., we recall that, at high \ac{AP} densities, we obtain small optimal beamwidths and thus high antenna directivity gains). Therefore, the SINR values of those \acp{UE} degrade the coverage to approximately 58\%; even so, this represents an improvement compared to the minimum distance strategy case.

In light of these results, for the \ac{UE} in pocket scenario, it is important to consider an association strategy that allows for the mitigation of  body shadowing effect, so as to provide satisfactory coverage. This is not the case for \acp{UE} in hand, as in this case the body shadowing effect on coverage is not as severe (as we can see by comparing the gains of the maximum received power association from Fig.~\ref{fig:cov_threshold_ap10_hand} and Fig.~\ref{fig:cov_threshold_ap10_pocket}).

\subsection{ASE vs. Coverage Trade-off}
\label{sec:tradeoff}

\begin{figure}
    \centering
    \includegraphics[width=\linewidth]{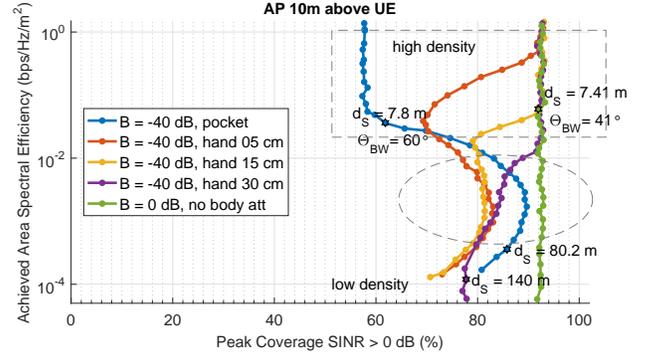}
    \caption{Trade-off between peak coverage (obtained for optimal beamwidth) and achieved ASE for different body attenuation scenarios. Inter-site distance increases from the upper points to the lower ones. The \ac{UE} associates with the \ac{AP} with maximum received power. \ac{UE} in pocket scenario needs larger beamwidths to achieve peak coverage. Points inside the rectangle provides best trade-off for \ac{UE} in hand scenario. Points inside the ellipse provide best coverage for \ac{UE} in pocket scenario and not lower ASE than inside the rectangle.}
    \label{fig:ase_cov_tradeoff_ap10_pocket_maxp}
\end{figure}

Finally, we analyzed the trade-off between the peak coverage and the achieved \ac{ASE} and its behaviour for different body shadowing scenarios and for different \ac{AP} densities. We investigate this trade-off for the SINR threshold of \unit[0]{dB}, which represents the \acp{UE} with higher receiver sensitivity; we focus on these \acp{UE} because they provide us with better insight on how the density and beamwidth settings affect network performance.
%
The results are shown in Fig.~\ref{fig:ase_cov_tradeoff_ap10_pocket_maxp}: each point of the curve corresponds to the optimal beamwidth $\theta_\mathrm{BW}$ for a given inter-site distance $d_S$ when using maximum power association. The lower points in the figure represent the ASE/coverage for larger $\theta_\mathrm{BW}$ and lower \ac{AP} density (longer $d_S$), while the upper points represent the ASE/coverage for smaller $\theta_\mathrm{BW}$ and higher \ac{AP} density (shorter $d_S$).
The main observation we make is that coverage and \ac{ASE} in the \ac{UE} in pocket scenario require different optimal beamwidth and \ac{AP} densities. For example, to optimize the coverage vs. \ac{ASE} trade-off in hand scenarios, the network should be designed to be dense and to use small beamwidths, i.e., choosing the points inside the gray rectangle of Fig.~\ref{fig:ase_cov_tradeoff_ap10_pocket_maxp}, where the coverage is around 90\% and the ASE is close to \unit[1]{bps/Hz/m$^2$}. However, the same configuration would yield poor coverage for \acp{UE} that are held in pocket.

A different design approach  aiming at coverage maximization for the \ac{UE} in pocket scenario requires deploying a sparser network with larger beamwidths (see points inside the ellipse Fig.~\ref{fig:ase_cov_tradeoff_ap10_pocket_maxp}). However, this design criterion is not optimal from the perspective of \ac{ASE};  as shown in the plot, \ac{ASE} suffers two orders of magnitude reduction as compared to the optimal value achievable with a denser network.

To summarize, we can optimize the design of indoor ceiling-mounted \ac{AP} mmWave networks either for the \ac{UE} in hand scenario or for the \ac{UE} in pocket scenario, but not both, as each scenario has different optimal configurations.

\section{Conclusion}

Herein we studied the performance effects of deployment densification of ceiling-mounted \ac{mmWave} access points with highly directional antennas over a confined area. We showed that, while being feasible, dense indoor \ac{mmWave} deployments have their intrinsic characteristics, which make it necessary for network designers to decide (and understand) what is their intended end user. First, the optimal choice of beamwidth maximizes either coverage or \ac{ASE}, but not both. Second, how this trade-off manifests itself will also depend on the human body shadowing scenario, i.e., the distance between the receiver and the potential obstruction, as the optimal choice of beamwidth and \ac{AP} density corresponds to the body blockage probability. 
As pointed out in Section \ref{sec:simulationsetup},
it is worth emphasizing that these findings are consistent across a range of area sizes and \ac{AP} heights relevant to the type of scenario we are considering.

Still more work is needed to understand how these trade-offs change when the \ac{mmWave} signals are scattered and reflected by the indoor environment. However, even the results we have so far can be readily used to inform the design of interference coordination techniques based on beam-steering, or new hand-off and cell association procedures that account for potential body shadowing of \ac{mmWave} signals.

\appendices

\section*{Acknowledgment}

This publication has emanated from the research conducted within the scope of \textit{NEMO (Enabling Cellular Networks to Exploit Millimeter-wave Opportunities)} project financially supported by the Science Foundation Ireland (SFI) under Grant No. 14/US/I3110 and with partial support of the European Regional Development Fund under Grant No. 13/RC/2077.

\section{}
\label{ap:simulationcodes}
All simulation scripts used to generate the presented results were written in MATLAB\textsuperscript{\textregistered} and can be cloned from the following repository: 

\textit{https://github.com/firyaguna/matlab-nemo-mmWave}

\ifCLASSOPTIONcaptionsoff
  \newpage
\fi



\bibliographystyle{IEEEtran}
\bibliography{main}
%



%





\end{document}